\documentclass[aps,prl,10pt,twocolumn,superscriptaddress,showpacs]{revtex4-1}
\usepackage{graphicx}
\usepackage{color}
\usepackage{bm}
\usepackage{multirow}

\begin{document}

\title{Discovery of low thermal conductivity compounds with first-principles anharmonic lattice dynamics calculations and Bayesian optimization}
\author{Atsuto \surname{Seko}}
\thanks{These authors contributed equally to this work.}
\affiliation{Department of Materials Science and Engineering, Kyoto University, Kyoto 606-8501, Japan}
\affiliation{Center for Elements Strategy Initiative for Structure Materials (ESISM), Kyoto University, Kyoto 606-8501, Japan}
\author{Atsushi \surname{Togo}}
\thanks{These authors contributed equally to this work.}
\affiliation{Center for Elements Strategy Initiative for Structure Materials (ESISM), Kyoto University, Kyoto 606-8501, Japan}
\author{Hiroyuki \surname{Hayashi}}
\affiliation{Department of Materials Science and Engineering, Kyoto University, Kyoto 606-8501, Japan}
\author{Koji \surname{Tsuda}}
\affiliation{Department of Computational Biology, Graduate School of Frontier Sciences, The University of Tokyo, Kashiwa 277-8561, Japan}
\author{Laurent \surname{Chaput}}
\affiliation{Institut Jean Lamour, UMR CNRS 7198, Universite de Lorraine, Boulevard des Aiguillettes, BP 70239, 54506 Vandoeuvre Les Nancy Cedex, France}
\author{Isao \surname{Tanaka}}
\email{tanaka@cms.mtl.kyoto-u.ac.jp}
\affiliation{Department of Materials Science and Engineering, Kyoto University, Kyoto 606-8501, Japan}
\affiliation{Center for Elements Strategy Initiative for Structure Materials (ESISM), Kyoto University, Kyoto 606-8501, Japan}
\affiliation{Nanostructures Research Laboratory, Japan Fine Ceramics Center, Nagoya 456-8587, Japan}

\date{\today}

\begin{abstract}
Compounds of low lattice thermal conductivity (LTC) are essential for seeking thermoelectric materials with high conversion efficiency.
Some strategies have been used to decrease LTC.
However, such trials have yielded successes only within a limited exploration space.
Here we report the virtual screening of a library containing 54,779 compounds.
Our strategy is to search the library through Bayesian optimization using for the initial data the LTC obtained from first-principles anharmonic lattice dynamics calculations for a set of 101 compounds.
We discovered 221 materials with very low LTC.
Two of them have even an electronic band gap $<$ 1 eV, what makes them exceptional candidates for thermoelectric applications.
In addition to those newly discovered thermoelectric materials, the present strategy is believed to be powerful for many other applications in which chemistry of materials are required to be optimized.
\end{abstract}

\maketitle

Thermoelectric generators are essential for utilizing otherwise waste heat.
Because of the technological importance, researchers have been seeking materials with high conversion efficiency for decades\cite{dresselhaus2007new,snyder2008complex,wan2010development,singh2008thermoelectrics}.
Compounds of low lattice thermal conductivity (LTC) are essential for this purpose.
Different strategies have been used to decrease LTC.
Recently, high throughput screening (HTS) of materials using materials database constructed by first principles calculations has been recognized as an efficient tool for accelerated materials discovery\cite{ceder2010opportunities,curtarolo2013high,Fujimura_2013_AENM:AENM201300060,nishijima2014accelerated,yu2013inverse}. 
Thanks to the recent progress of computational power and techniques, a large set of first principles calculations can be performed with the accuracy comparable to experiments.
This is a straightforward strategy when both of the following conditions are satisfied: 1) the target physical property can be accurately computed by first principles methods.
2) The exploration space is well defined and not too large to compute the target physical property exhaustively in the space. 

In order to evaluate LTC with the accuracy comparable to experimental data, however, we need to develop a method that is far beyond the ordinary density functional theory (DFT) calculations.
Since we need to treat multiple interactions among phonons, or anharmonic lattice dynamics, the computational cost is many orders of magnitudes higher than the ordinary DFT calculations.
Such expensive calculations are practically possible only for a small number of simple compounds.
HTS of a large DFT database of LTC is not a realistic approach unless the exploration space is narrowly confined.
In the year 2014, Carrete and coworkers concentrated their efforts to search low LTC materials within half-Heusler compounds\cite{carrete2014finding}.
They made HTS of wide variety of half-Heusler compounds by examination of thermodynamical stability via DFT results.
Then LTC was estimated either by full first principles calculations or by a machine-learning algorithm for a selected small number of compounds.
HTS of low LTC using a quasiharmonic Debye model was also reported in 2014\cite{toher2014high}.
Efficient prediction of LTC through compressive sensing of lattice dynamics was recently demonstrated\cite{zhou2014lattice}.
Development of such new methods would bring accelerated discovery of new materials in the future.

In the present study, we do not want to restrict the exploration space by empirical knowledge, for example, by crystal structure.
We firstly evaluated LTC of 101 compounds with three prototype structures, i.e., rocksalt, zincblende and wurtzite-type structures, by first-principles anharmonic lattice dynamics calculations and solving Boltzmann transport equation with the single-mode relaxation-time approximation\cite{togo2015distributions,togo2008first}.
Then the results are used to construct a model for making ``virtual screening" of many compounds in a library with a diversity of structures and chemical compositions employing Bayesian optimization procedure.
For the Bayesian optimization, predictors are determined by kriging method to find the lowest LTC compound among the 101 first principles data.
The highly ranked compounds are supplied to first principles LTC calculations to verify the result of the screening. 

Computational procedure of LTC is described in detail elsewhere\cite{togo2015distributions}.
LTCs were calculated from phonon lifetimes, group velocities, and mode-heat capacities solving the phonon Boltzmann transport equation within the relaxation time approximation.
The phonon properties were calculated from the force constants.
We employed first-principles calculation to obtain second-order force constants (FC2) and third-order force constants (FC3) with the supercell and finite displacement approaches.
Phonopy code was used for these phonon calculations\cite{togo2008first}. 
Finite displacements of 0.03 \AA\ were systematically introduced to perfect supercells to fill up all elements of force constant tensor elements among atoms in the supercells. 
The Brillouin zone integration for the phonon lifetime calculation was performed by the linear tetrahedron method. 

\begin{figure*}[tbp]
\begin{center}
\includegraphics[width=0.7\linewidth,clip]{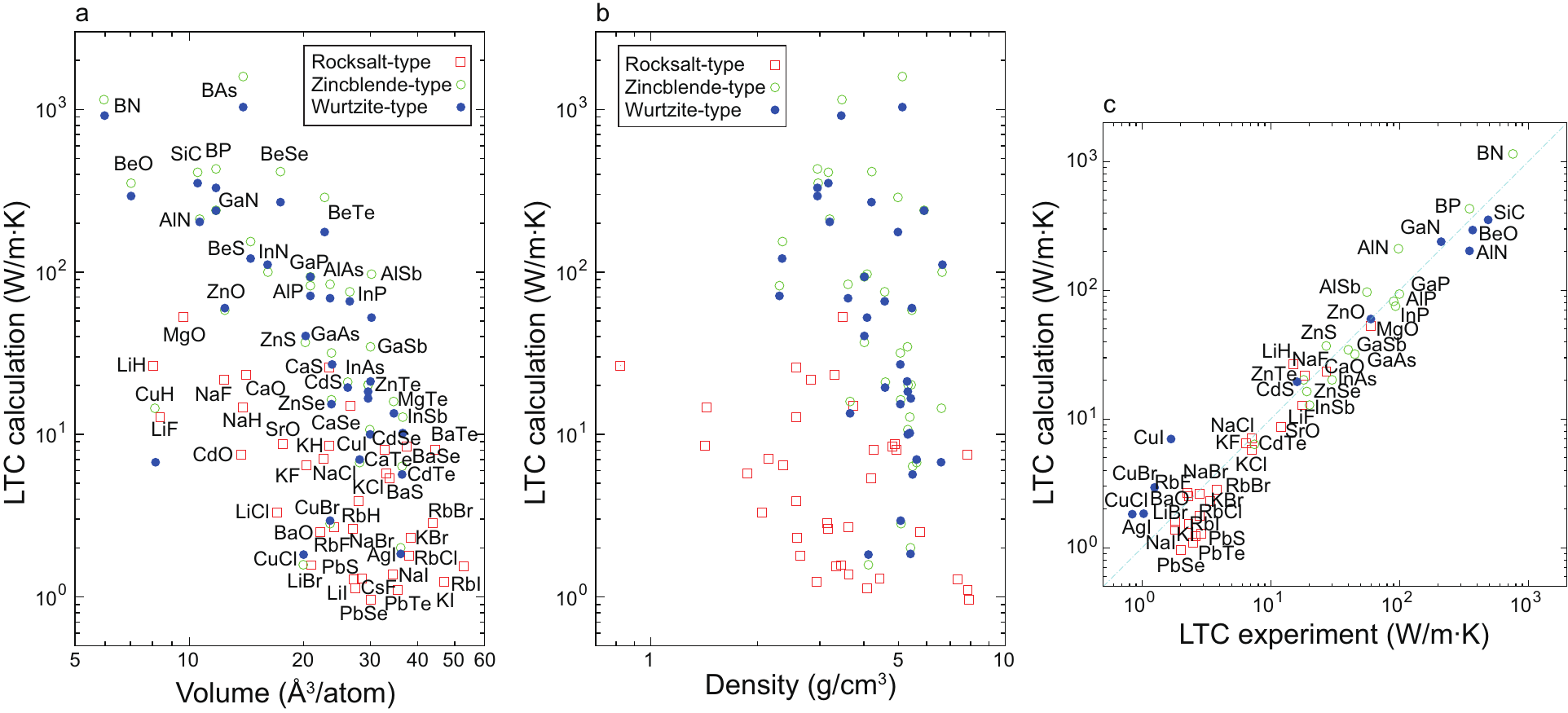} 
\caption{
LTC calculated from first principles for 101 compounds along with (a) volume, $V$, and (b) density, $\rho$.
(c) Experimental LTC data are shown for comparison when experimental LTCs are available.
}
\label{thermal:fig1}
\end{center}
\end{figure*}

For the first principles calculations, we employed the plane-wave basis projector augmented wave (PAW) method\cite{PAW1} in the framework of DFT and the generalized gradient approximation of the Perdew-Burke-Ernzerhof (PBE) form\cite{GGA:PBE96} as implemented in the VASP code\cite{VASP1,VASP2,PAW2}.
Much more attention for the convergence of DFT calculations should be paid in the phonon calculations as compared to the ordinary first principles calculations with respect to the k-point mesh, plane wave energy cutoff and tolerances of energy, residual force and stress. 
The size of the supercell was chosen by observing the convergence of phonon properties by changing the supercell size.
Low LTC crystals are generally more anharmonic and the atomic interaction range is considered relatively large.
LTC calculations of the highly ranked compounds required larger supercells than those for ordinary crystals with smaller anharmonicity.
The plane wave energy cutoff was chosen to be at least 20\% higher than the recommended values in the PAW dataset.
Total energies were minimized until the energy convergences became less than $10^{-8}$ eV. 

Results of first principles LTC of 101 compounds are shown with crystalline volume per atom, $V$, and density, $\rho$, in Figs. \ref{thermal:fig1} (a) and (b).
Among 101 compounds, PbSe with the rocksalt structure show the lowest LTC, 0.9 W/m·K (@300 K).
It is in the similar trend as the recent report showing low LTC for lead- and tin-chalcogenides\cite{lee2014resonant,parker2013high,zhao2014ultralow,skelton2014thermal}.
The computed results are compared with available experimental data in Fig. \ref{thermal:fig1} (c).
Satisfactory agreements between experimental and computed results are evident in Fig. \ref{thermal:fig1} (c), demonstrating the usefulness of the first principles LTC data for further studies.
A phenomenological relationship has been proposed that $\log \kappa_L$ is proportional to $\log V$\cite{slack1979thermal}.
Although qualitative correlation can be seen between our LTC and $V$, it is difficult to predict LTC quantitatively, hence to discover new compounds with low LTC, only from the phenomenological relationship.
It can be noted that the dependence on $V$ is remarkably different between rocksalt type and zincblende or wurtzite type compounds, while zincblende and wurtzite type compounds show similar LTC when the chemical compositions are the same. 

The 101 first principles LTC data are then used to make a model for the prediction of LTC of compounds within a library on the basis of the Bayesian optimization.
For the purpose of the prediction, it is preferable to select ``good" predictors.
Our rule of thumb is as follows: 1) whenever experts' knowledge is available as a physical or phenomenological rule, it should be examined as the first step.
2) Predictors may be better included in a library or those easily made by combining the physical quantities in a library.
Alternatively, the predictors may be easily computed by DFT calculations.
3) High efficiency for the Bayesian optimization procedure needs to be examined.

On the basis of these ideas, we firstly determine predictors for the Bayesian optimization procedure to find the lowest LTC compound among the 101 first principles LTC data.
We adopt kriging method based on the Gaussian process regression (GPR)\cite{Rasmussen2006Gaussian,seko2014machine} of LTC simply using two physical quantities, $V$ and $\rho$, as predictors.
These quantities are available in most of the experimental or computational crystal structure database, such as ICSD\cite{bergerhoff1987crystal}, Atomwork\cite{xu2011inorganic}, Materials Project Database (MPD)\cite{jain2013commentary}, and aflowlib\cite{curtarolo2012aflowlib}.
Although a phenomenological relationship has been proposed between $\log \kappa_L$ and $V$\cite{slack1979thermal}, the correlation between them is not so high.
The correlation between $\log \kappa_L$ and $\rho$ is even worse.

We start from an observed data set of 5 compounds that is randomly chosen from 101 compounds. 
In the kriging, a compound with maximum probability of improvement among the remaining data is searched, namely a compound with the highest Z-score derived from GPR.
The compound is included into the observed data set and then another compound with maximum probability of improvement is searched.
Both the kriging and random searches are repeated fifty times and the average number of observed compounds required for finding the compound with the lowest LTC is examined.


When $- \log \kappa_L$ is expressed as $f$, Z-score for a compound with predictors $\bm{x}^*$ is defined as 
\begin{equation}
Z(\bm{x}^*) = \left[ f(\bm{x}^*) - f_{\rm best} \right] / \sqrt{v(\bm{x}^*)}
\end{equation}
where $f(\bm{x}^*)$ and $v(\bm{x}^*)$ denote the predicted value of $- \log \kappa_L$ and its prediction variance at a point expressed by predictors $\bm{x}^*$, respectively. 
$v(\bm{x}^*)$ is expected to be small for compounds near the observed data, while it can be large for compounds far from the observed data.
$f_{\rm best}$ denotes the lowest LTC value among ``observed" compounds, which is updated at each kriging step.
Z-score that is evaluated by dividing $\left[ f(\bm{x}^*) - f_{\rm best} \right]$ by the square root of the prediction variance, $\sqrt{v(\bm{x}^*)}$ tends to select candidates with maximum probability of improvement\cite{jones01}.
Here the prediction and its variance are described using the Gaussian kernel function.
Therefore, our GPR has two free parameters, i.e. variances of Gaussian kernel and prior distribution. 
Here, they are given as 20 and 0.1, respectively. 

\begin{figure}[tbp]
\begin{center}
\includegraphics[width=\linewidth,clip]{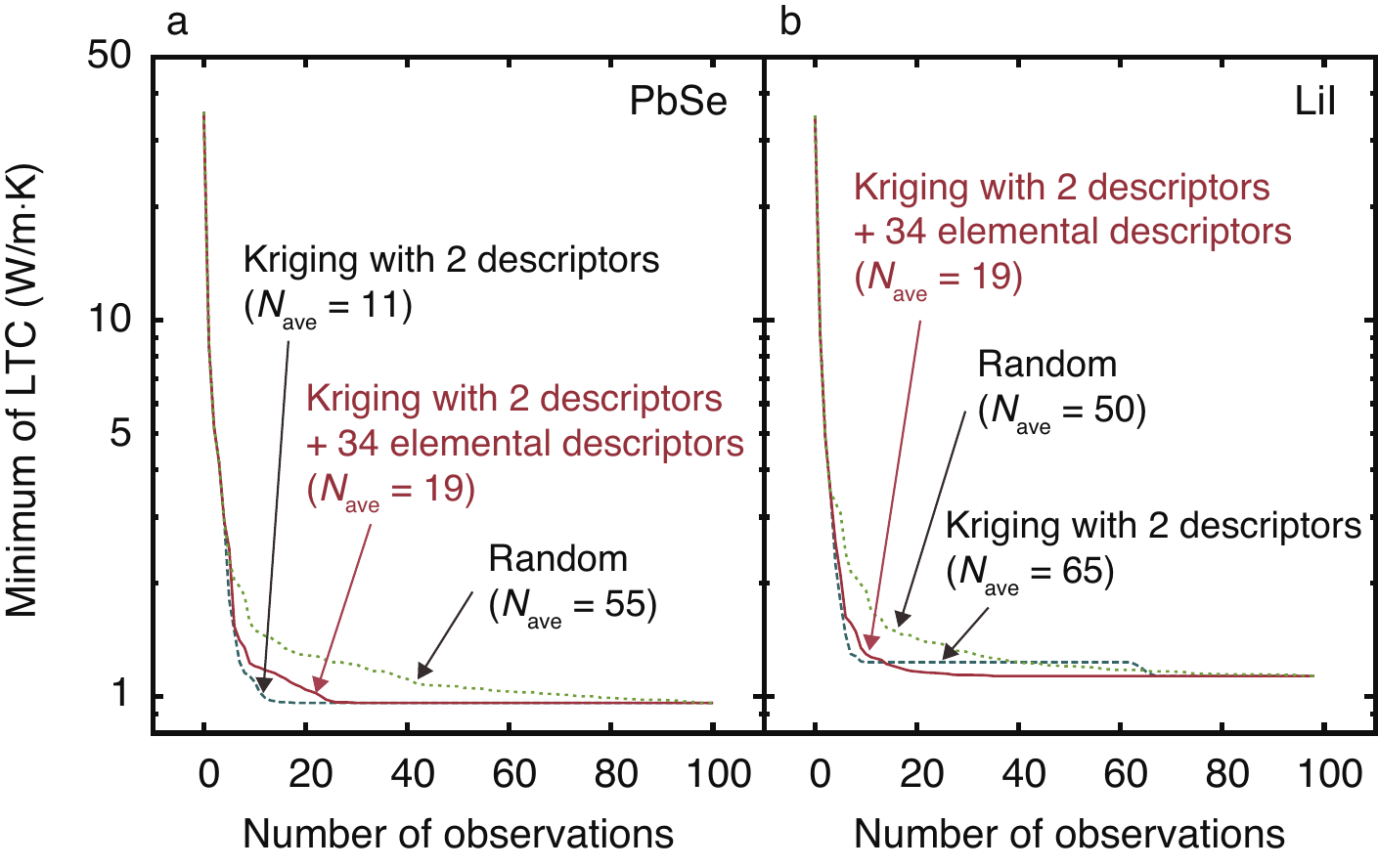}
\caption{
Lowest LTC values at each iteration in kriging search for finding (a) PbSe and (b) LiI. Those by random searches are also shown for comparison. When performing a kriging search for finding LiI, PbSe and PbTe are intentionally omitted and the rest of 99 compounds are used.
}
\label{thermal:fig2}
\end{center}
\end{figure}

Figure \ref{thermal:fig2} (a) shows the result of the kriging search in comparison to the random search of the lowest LTC compounds within the 101 compounds.
The average numbers of compounds required for the optimization using the kriging and random searches, $N_{\rm ave}$, are 11 and 55, respectively.
The compound with the lowest LTC among the 101 compounds, i.e., rocksalt PbSe, can be found much more efficiently using the kriging technique and only with two variables, $V$ and $\rho$.
However, we realize that the kriging only with these two variables is not a robust way for finding the lowest LTC.
As an example, Fig. \ref{thermal:fig2} (b) shows the result of the kriging search using the dataset after intentionally removing 1st and 2nd lowest LTC compounds, i.e., rocksalt PbSe and PbTe, from the 101 compounds.
Then rocksalt LiI should be the right answer of the optimization.
However, $N_{\rm ave}$ is 65 for finding LiI using the kriging only with $V$ and $\rho$, which is larger than that of the random search, $N_{\rm ave} = 50$.
The delay of the optimization should originate from the fact that LiI is an outlier when LTC is modeled only with $V$ and $\rho$.
Such outlier compounds with low LTC are difficult to find only with $V$ and $\rho$.

In order to overcome the outlier problem, we add predictors about constituent chemical elements.
There are many choices for such variables: 
They are, for example, electronegativity, atomic radius, ionization energy, etc\cite{seko2014machine}.
Here, we newly introduced ``elemental descriptors", which is a set of binary digits representing the presence of chemical elements.
Since the 101 LTC data is composed of 34 kinds of elements, we use 34 elemental descriptors. 
Results of the kriging are shown in Figs. \ref{thermal:fig2} (a) and (b) with 34 elemental descriptors on top of $V$ and $\rho$. 
In both cases, the compound of the lowest LTC is found with $N_{\rm ave} = 19$.
The use of the elemental descriptors is found to improve the robustness of the efficient search. 

As described in the Supplemental Material (SM), better correlations with LTC can be found for parameters that are obtained from phonon density of states.
However, we do not use such phonon parameters as predictors in the present study, because there is no data library available for such phonon parameters for a wide range of compounds.
Hereafter, we show results only with the predictor set composed of 34 elemental descriptors on top of $V$ and $\rho$.

\begin{figure}[tbp]
\begin{center}
\includegraphics[width=\linewidth,clip]{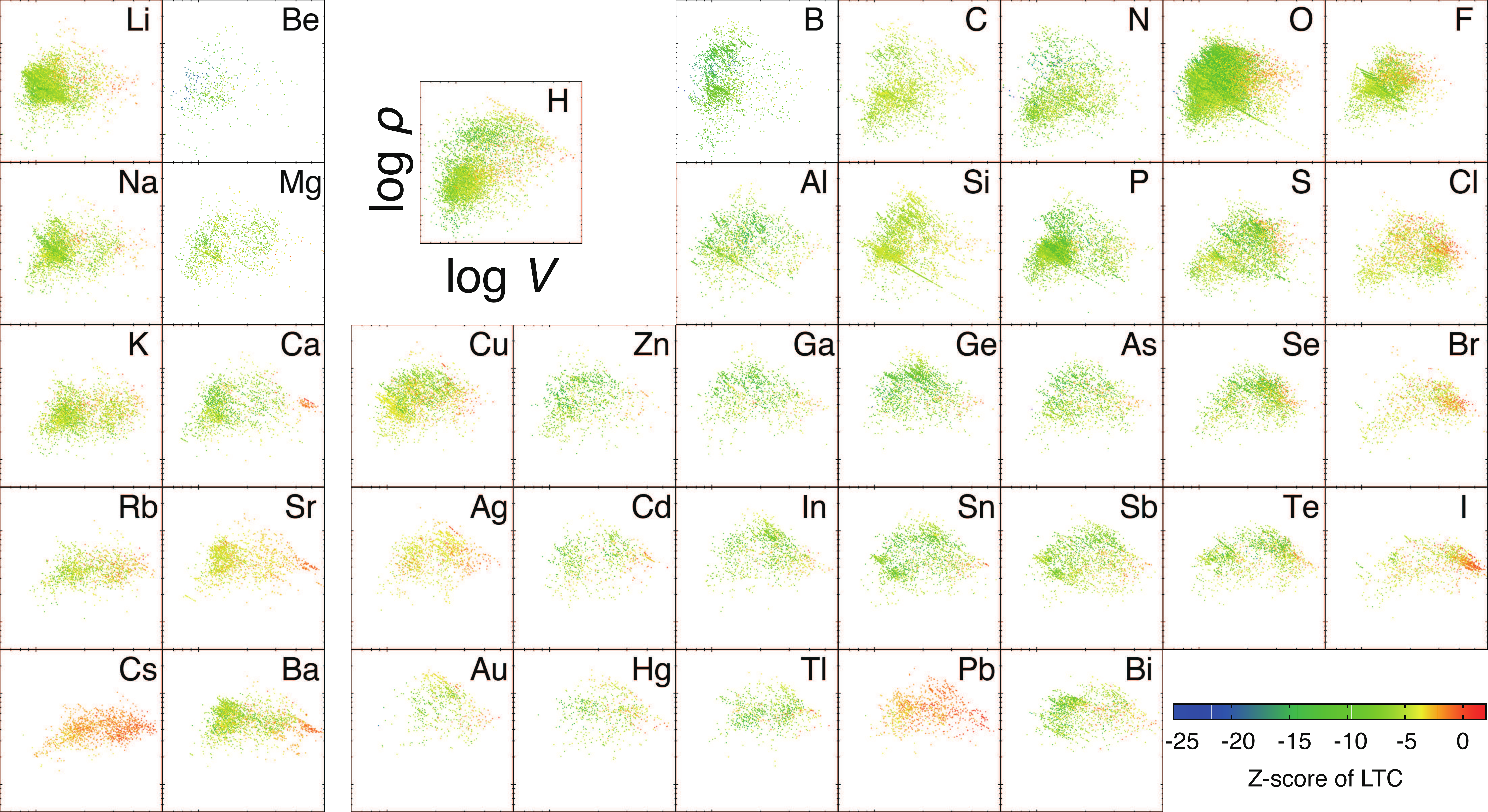} 
\caption{
Dependence of Z-score on constituent elements for compounds in the MPD library. The magnitude of Z-score is shown by colors along with volume, $V$, and density, $\rho$, for each element.
}
\label{thermal:fig3}
\end{center}
\end{figure}

Screening for low LTC compounds over compounds in a large library is carried out using a GPR prediction model.
Such a screening based on a prediction model is called ``virtual screening" in biomedical communities\cite{kitchen2004docking}.
For the virtual screening, we adopt all 54,779 compounds in MPD library\cite{jain2013commentary,ong2013python}, which is composed of most of crystal structure data available in ICSD\cite{bergerhoff1987crystal}.
On the basis of the GPR prediction model made by $V$, $\rho$ and 34 elemental descriptors for the 101 LTC data, a ranking for low LTC compounds is made according to the Z-score of the 54,779 compounds. 

Figure \ref{thermal:fig3} shows the distribution of Z-scores for the 54,779 compounds along with $V$ and $\rho$.
The magnitude of Z-score is plotted in panels corresponding to constituent elements.
(Transition metal and other elements are shown in SM.) 
The Z-score is relative to rocksalt PbSe showing the lowest LTC among 101 compounds. 
Among 54,779 compounds, 221 compounds show positive Z-score, which are expected to have lower LTC than that of rocksalt PbSe, i.e., $<$ 0.9 W/mK (@300 K). 
They are highlighted by red dots. 
They are widely distributed in $V-\rho$ space; which means it is difficult to pick them up without performing the Bayesian optimization with elemental descriptors. 
The Z-score is widely distributed for light elements such as Li, N, O and F.
This implies that the presence of such light elements by itself have little effects on lowering the LTC. 
When such light elements form a compound with heavy elements, the compound tends to show high Z-score. 
It is also noteworthy that many compounds composed of some light elements such as Be and B tend to show high LTC. 

Special features are recognized for Pb, Cs, I, Br and Cl.
Many compounds composed of these elements exhibit high Z-score.
(The number of compounds with positive Z-score is shown in SM.) 
Most of compounds showing positive Z-score have any of atomic combinations of these five elements.
On the other hand, elements in the Periodic table neighboring to these five elements do not show analogous trends. 
For example, compounds with high Z-scores are rarely found for Tl and Bi, which are neighboring to Pb. 
This may sound odd since Bi$_2$Te$_3$ is a famous thermoelectric compound. 
This may be ascribed to our selection of the training dataset composed only of AB compounds with 34 elements and three kinds of simple crystal structures.
In other words, the training dataset is somehow ``biased". 
This is unavoidable at the moment since the first-principles LTC calculations are still too expensive to obtain sufficiently unbiased training dataset with a large enough number of data to cover the diversity of chemical composition and crystal structures.
Nevertheless, the ``biased" training dataset will be verified to be useful for finding low LTC materials. 
Because of the use of the ``biased" training dataset, we may not be able to discover all of the low LTC materials in the library. 
However, we can discover at least a part of them.   

Verification process for the candidates of low LTC compounds after the virtual screening is one of the most important steps to ``discover" low LTC compounds.
First principles LTCs are evaluated for the top 8 compounds after the virtual screening.
All of them are considered to form ordered structures.
LTC calculation was unsuccessful for Pb$_2$RbBr$_5$ due to the presence of imaginary phonon modes within the supercell used in the present study.
Z-scores and first principles LTC of the rest of the compounds are listed in Table 1. 
All of top 5 compounds show LTC of $<$ 0.2 W/mK (@300 K), which are much lower than that of the rocksalt PbSe, i.e., 0.9 W/mK (@300 K).
This confirms the powerfulness of the present GPR prediction model for efficiently discovering low LTC compounds.
Crystal structures of highly ranked compounds, PbRbI$_3$, PbIBr, PbRb$_4$Br$_6$ and PbI$_2$ ($P6_3mc$) are shown in SM. 
PbICl and PbClBr have the same crystal structures as PbIBr. PbI$_2$ ($R\overline{3}m$) and PbI$_2$ ($P6_3mc$) are different only in their stacking sequences. 
All of these compounds contain either six-fold or eight-fold coordinated Pb by halogen ions, and are of stoichiometric chemical composition when Pb is divalent. 
\begin{figure}[tbp]
\begin{center}
\includegraphics[width=0.8\linewidth,clip]{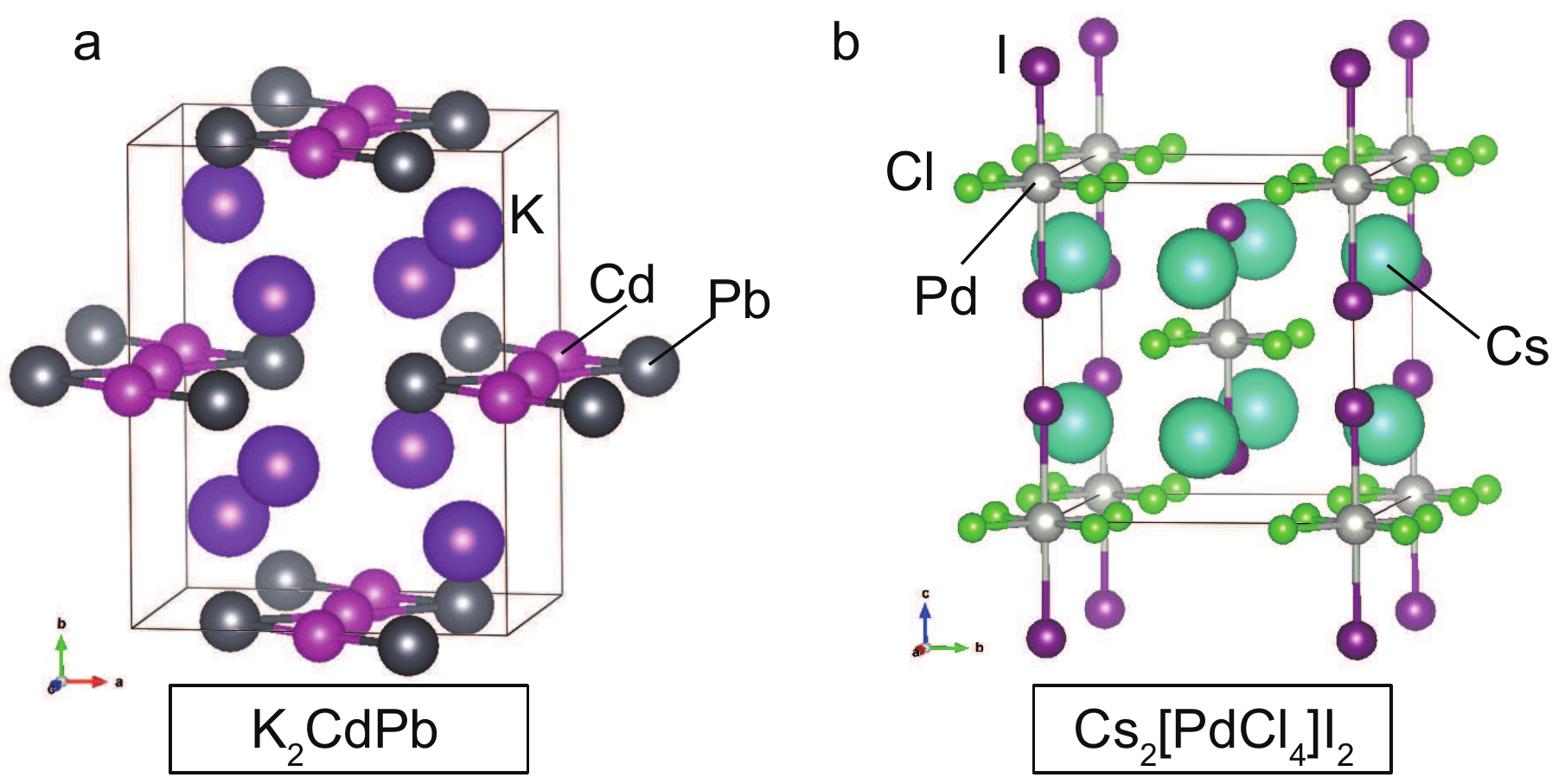} 
\caption{
Crystal structures of K$_2$CdPb and Cs$_2$[PdCl$_4$]I$_2$ predicted to show low LTC of $<$ 0.5 W/mK (@300 K) and narrow band gap of $<$ 1 eV.
}
\label{thermal:fig4}
\end{center}
\end{figure}

\begin{table}[tbp]
\caption{
First principles LTCs and Z-scores for highly ranked compounds by the virtual screening. 
Band gaps by DFT-PBE are taken from MPD library\cite{jain2013commentary,ong2013python}.
}
\label{thermal:table1}
\begin{ruledtabular}
\begin{tabular}{cccccc}
\multirow{2}{*}{Ranking} & \multirow{2}{*}{Z-score} & \multirow{2}{*}{Formula} & Space & LTC & Band  \\
&  & & group & (W/mK) & gap (eV) \\
\hline
1   &  1.90  &  PbRbI$_3$        & $Pnma $   &  0.10    &  2.46  \\
2   &  1.76  &  PbIBr            & $Pnma $   &  0.13    &  2.56  \\
3   &  1.56  &  PbRb$_4$Br$_6$   & $R\overline{3}c $   &  0.08    &  3.90  \\
4   &  1.56  &  PbICl            & $Pnma $   &  0.18    &  2.72  \\
5   &  1.56  &  PbClBr           & $Pnma $   &  0.09    &  3.44  \\
7   &  1.44  &  PbI$_2$          & $R\overline{3}m $   &  0.29    &  2.42  \\
8   &  1.43  &  PbI$_2$          & $P6_3mc$   &  0.29    &  2.45  \\
\hline
121 &  0.39  &  K$_2$CdPb        & $Ama2 $   &  0.45    &  0.18  \\
144 &  0.29  &  Cs$_2$[PdCl$_4$]I$_2$  & $I4/mmm$  &  0.31    &  0.88 
\end{tabular}
\end{ruledtabular}
\end{table}

When such LTC materials are considered for thermoelectric applications, properties related to electronic structures, namely electronic contribution of the thermal conductivity, electrical conductivity and Seebeck coefficient should also be optimized. 
Although they can be tuned by elemental doping, the band gap, $E_g$, should be a simple measure of the electronic structure and allows to discriminate in a simple way between materials that can be good thermoelectrics or not. 
All of 221 compounds showing positive Z-score are listed in SM together with $E_g$ (DFT-PBE) given in the MPD library. 
Among them only 19 compounds satisfy 0.1 $ < E_g < $ 1.0 eV. 
First principles LTCs are evaluated for them. 
Crystal structures and LTC for two of them are shown in Fig. \ref{thermal:fig4} and Table \ref{thermal:table1}. 
Both of K$_2$CdPb and Cs$_2$[PdCl$_4$]I$_2$ are predicted to exhibit LTC of less than 0.5 W/mK (@300 K) together with band gap of smaller than 1 eV.
The discovery of such compounds may open a gate toward designing new thermoelectric materials with exceptionally high figure of merit.

In this study, we first report the theoretical LTC of 101 compounds by first-principles anharmonic lattice dynamics calculations.
Using these data, the virtual screening of a library containing 54,779 compounds is performed by Bayesian optimization using kriging method based on the Gaussian process regressions.
221 materials with very low LTC are found from this screening.
A final filtering of those low LTC compounds is made using the electronic band gap, which is a measure to discriminate in a simple way between materials that can be good thermoelectrics or not.
Two compounds with low LTC of $<$ 0.5 W/m·K (@300K) and narrow band gap of $<$ 1 eV are thus discovered, which may open a gate toward designing new thermoelectric materials with exceptionally high figure of merit.
The present method should be useful for searching materials for many different applications in which chemistry of materials are required to be optimized. 

This work was supported by Grant-in-Aid for Scientific Research (A) and Grant-in-Aid for Scientific Research on Innovative Areas ``Nano Informatics" (Grant No. 25106005) from the Japan Society for the Promotion of Science (JSPS).
\bibliography{thermal}

\end{document}